\documentclass[preprint2]{aastex}

\shorttitle{BAL Frequency}
\shortauthors{Hewett and Foltz}
\begin{document}
\def\kms{{\rm\,km\,s^{-1}}}
\title{THE FREQUENCY AND RADIO PROPERTIES OF BROAD ABSORPTION LINE 
QUASARS\altaffilmark{1}
\altaffiltext{1}{Observations reported here were obtained in part at
the MMT Observatory, a joint facility of the Smithsonian Institution
and the University of Arizona}}

\author{Paul C. Hewett}
\affil{Institute of Astronomy, Madingley Rd, Cambridge, CB3 0HA, United Kingdom}
\email{phewett@ast.cam.ac.uk}

\and

\author{Craig B. Foltz}
\affil{MMT Observatory, University of Arizona, Tucson, AZ 85721}
\email{cfoltz@as.arizona.edu}

\begin{abstract}

A sample of 67 Broad Absorption Line quasars (BALQSOs) from the Large Bright Quasar
Survey (LBQS) is used to estimate the observed and intrinsic fraction of BAL
quasars in optically--selected samples at intermediate ($B_J \simeq 18.5$)
magnitudes.  The observed BALQSO fraction in the redshift range $1.5 \le z
\le 3.0$ is $15\pm3\%$.  A well--determined, empirical, $k$--correction, to
allow for the differences in the spectral energy distributions of non--BALQSOs and
BALQSOs shortward of $\simeq 2100$\AA \ in the restframe, is applied to the
sample.  The result is an estimate of the intrinsic fraction of BALQSOs, in
the redshift range $1.5 \le z \le 3.0$, of $22\pm4\%$.  This value is twice that
commonly cited for the occurrence of BALQSOs in optically--selected samples and the
figure is in reasonable agreement with that from a preliminary analysis of the
SDSS Early Data Release.  The fraction of BALQSOs predicted to be present in
an optical survey with flux limits equivalent to that of the FIRST Bright Quasar
Survey (FBQS) is shown to be $\simeq 20\%$.  The BALQSO fractions derived
from the FBQS and the LBQS suggest that optically--bright BALQSOs are 
half as likely as non-BALQSOs to be detectable as 
$S_{1.4{\rm GHz}}\ga 1\,$mJy radio sources.

\end{abstract}


\section{INTRODUCTION}

Virtually every paper written in the last decade describing the phenomenology of
the broad absorption line quasars (BALQSOs) contains the word ``enigmatic''.
Long thought to be a rare sub--class of quasars relegated to an esoteric
sub--discipline of active galactic nuclei (AGN) studies, these objects have
garnered new attention in recent years.  It has become clear that the BALQSOs,
while a minority class of quasars, are not rare.  Furthermore, the evidence is
strong that the absorbing gas responsible for the prominent, blueward--displaced
absorption troughs from species covering a range of ionization from \ion {Mg}
{2} and \ion {Fe} {2} through \ion {O} {6} may also be responsible for the
``warm absorbers'' seen at X-ray wavelengths (Brandt, Laor \& Wills 2000).  
The BAL phenomenon is also almost
certainly related to the ultraviolet absorption seen at low ejection velocities
in a number of low luminosity AGN and to the ``associated absorbers'' seen in
quasars at higher redshifts.  A summary of the state of understanding mass
outflows in AGN and the links between the BAL phenomena and other manifestations
of outflows can be found in contributions included in Crenshaw, Kramer \& George
(2002).

While some progress has been made in understanding the link between the
presence of broad absorption and the properties of quasar spectral energy
distributions at other wavelengths, there is still no consensus concerning
the origin and acceleration mechanism of the absorbing gas in BALQSOs.
Indeed, the fundamental relation between BALQSOs and the quasar population
as a whole remains open.  Two broad classes of models have been proposed;
what can be termed the ``unified'' and ``evolutionary'' schemes.  In the
former, whether a quasar is observed to have broad absorption troughs depends
on our viewing angle to the quasar with respect to some preferred axis
(Weymann et al. 1991; hereafter WMFH). In the evolutionary scheme, the broad
absorption troughs are produced during some specific period in the evolution of
the quasar, perhaps as it transforms itself from a fully--enshrouded object
with a large infrared luminosity, through a BAL phase, into a normal quasar
(e.g. Briggs, Turnshek \& Wolfe 1984; Sanders 2002).

Independent of which general model holds (or if indeed neither obtains), the
frequency of occurrence of the BAL phenomenon provides direct constraints on
key physical parameters (e.g. Morris 1988).  In the unified--scheme, the
frequency bears on the covering factor of the BAL gas and, in the
evolutionary--scheme, the frequency is related to the time spent in the BAL
quasar phase relative to that spent as a normal quasar.  Determining the frequency
of the BAL phenomena as a function of redshift and strength of the absorption
features would provide further constraints on the competing models.
Notwithstanding the importance of establishing the frequency of the BAL
phenomenon, there are very few such quantitative estimates in the literature.
Few estimates that have been published contain both a detailed description of
the calculation and are based on well--defined samples of quasars with
numbers of objects sufficient to reduce the errors to an interesting level.
For example, Foltz et al.  (1992) provide few details of their calculation,
Chartas (2000) employs a novel method but has very few objects and the recent
determination by Tolea, Krolik \& Tsvetanov (2002), based on the Sloan
Digital Sky Survey (SDSS) Early Data Release (EDR), uses a large number of
BALQSOs but the selection function, for quasars with different spectral
energy distributions included in the EDR, is yet to be determined.

The difficulty in determining the frequency of BALQSOs is tied to their
selection via the effect of the absorbing material along the line--of--sight on the
spectral energy distribution (SED) of the quasars.  Over extended ranges in
frequency, such as the X--ray portion of the spectrum, BALQSOs may be
under--represented in flux--limited samples due to the effects of the
absorbing column.  Within narrower frequency intervals, particularly in the
rest--frame ultraviolet and optical portions of the SED, the extreme
absorption equivalent widths of the BAL troughs can suppress the broad--band
magnitudes well below those of a non--BALQSO of comparable luminosity and
redshift to the extent that the BALQSO is excluded from a magnitude--limited
sample.  This effect is strongly dependent on both redshift, since most of
the strong troughs occur in the rest--frame ultraviolet, and on ionization
level, where some of the most extreme low--ionization BALQSOs show very
strong depressions shortward of 2800\AA.

It is still common practice to quote a figure of $\sim 10\%$ for the observed
frequency of BALQSOs in the quasar population.  The Large Bright Quasar
Survey (Hewett, Foltz \& Chaffee 1995, 2001) is cited as one of the main surveys
from which the $\sim 10\%$ is derived (e.g.  Stocke et al.  1992) and current
observations of samples of BALQSOs using {\it Chandra} (Green et al.  2001;
Gallagher et al.  2002) are based on samples drawn from the LBQS.  Recently,
Becker et al.  (2000) used a figure of $10\%$ for the BALQSO frequency in
the LBQS, in conjunction with quasars from the FIRST Bright Quasar Survey
(FBQS), to conclude that BALQSOs were in fact significantly more common in
the FBQS than in optically--selected samples.  This latter conclusion is at first
sight inconsistent with the long--standing belief that BALQSOs are extremely
rare amongst the population of quasars that are also luminous radio sources
(Stocke et al.  1992).

Weymann (2002) has stressed the need to ensure that investigation of outflow
phenomena in quasars does not become handicapped by placing great significance
on precise definitions of classes of object.  Rather, attention
should be paid to trends that may provide insight into the underlying physics.
Indeed, Weymann cites the potential limitations inherent in the original
quantitative definition of the strength of BALs in spectra, the BALnicity Index
(WMFH), given the subsequent recognition that outflows may also be responsible
for many narrow absorption lines with much lower ejection velocities.  However,
clearly defined and reproducible schemes for the identification of BALQSO
samples are a prerequisite if quantitative comparisons between the incidence
of BALQSOs as a function of the properties of quasar SEDs are to be performed.
For example, given the availability of spectra from the SDSS, Hall et al.{}
(2002) have sensibly extended the original definition of the WMFH BALnicity Index
to incorporate BAL troughs associated with low--ionization species, spectacular
examples of which are included among the SDSS quasar sample.

Substantial progress in understanding the relation between BALQSOs and the
quasar population as a whole has already been made given the availability of
large samples of quasars from the SDSS (Hall et al.  2002; Tolea et al.  2002)
and the FBQS (Becker et al.  2000).  Further progress can be expected from these
and other surveys, particularly those undertaken at near--infrared wavelengths,
in the next few years.  However, the lack of a well--determined fraction of BAL
quasars from existing surveys, the continued use of samples from the
Palomar--Green (Schmidt \& Green 1983) and LBQS surveys for follow--up studies
at other wavelengths (e.g.  Green et al.  2001), combined with the seemingly
contradictory conclusions regarding the radio--properties of BALQSOs argue
for a reconsideration of the incidence of BALQSOs derived from apparently
bright, optically--selected quasar surveys.

In \S2 a sample of 67 candidate BALQSOs from the full LBQS sample is
presented.  A subset of 42 objects form a well--defined sample with redshifts
$1.5 \le z \le 3.0$.  Section 3 presents calculations of the observed fraction
of BALQSOs in the LBQS.  A two--dimensional BALQSO $k$--correction is
developed to account for the under--representation of BALQSOs in the
flux--limited LBQS and an estimate of the intrinsic fraction of BALQSOs is
presented.  Synthetic photometry applied to composite quasar spectra is used in
\S4 to calculate the fraction of optically--selected quasars predicted to be
present in a sample of quasars with the same photometric selection criteria as
the FBQS.  Very different conclusions from those of Becker et al.  (2000),
regarding the probability a BALQSO is detected in the FIRST survey, are
reached.  The paper concludes with a discussion of the constraints on the
demographics of BALQSOs from the LBQS, FBQS and SDSS surveys in \S 5.

\section{THE BROAD ABSORPTION LINE QUASAR SAMPLE}

The incidence of BALQSOs in the LBQS has been determined from a
re--examination of the discovery spectra (Foltz et al. 1987, Foltz et
al. 1989, Hewett et al. 1991, Chaffee et al. 1991, Morris et al. 1991,
Hewett et al.  2001) of the current LBQS sample, comprising the 1055
objects from Hewett et al. (1995) and the 12 additional
objects listed in Table 1 of Hewett et al. (2001). In
practice, the objects of interest are predominantly those flagged in the
discovery papers as ``b'', ``b?'' or ``b??'', corresponding to
definite, probable or possible BALQSOs. A significant number of these
objects possess additional spectra of higher quality from the
investigations of WMFH and Korista et al. (1993).

Classification of a quasar as a BALQSO was based on the BALnicity index
of WMFH applied to the \ion{C}{4} $\lambda$1549 emission line.  The
BALnicity index attempts to characterize the absorption by measuring
essentially the equivalent width of the absorption feature, expressed
in $\kms$, with the additional constraints that the absorption must be
contiguous over $2000\kms$ and must lie more than $3000\kms$ blueward
of the broad emission line redshift.  The \ion{C}{4} $\lambda$1549
emission line and any associated absorption lie within the atmospheric
window only for redshifts $z \ga 1.3$.  Furthermore, to determine a
BALnicity for an object it is necessary to be able to define a reliable
continuum level.  For quasars with redshifts $z \sim 1.4$ the proximity
of \ion{C}{4} $\lambda$1549 absorption to the atmospheric edge,
combined with the only moderate quality of some of the spectra, means
that a reliable continuum level could not always be defined, precluding
the measurement of the BALnicity for a small number of objects.  There
are six such quasars included in Table 1, five of which are assigned to
the ``possible'' BAL category.  Only one candidate BALQSO, B2210-1751,
among the sub--sample of 42 candidate BALQSOs used in subsequent
sections is without a measured BALnicity Index.  A modest value of
BALnicity$=1000\kms$ has been adopted for B2210-1751 for the purposes
of the calculations in \S 3 onwards.

For quasars with redshifts $z < 1.2$, identification of (low--ionization) BALQSOs,
was based on the presence of absorption associated with the \ion{Mg}{2}
$\lambda$2798 emission line.  The WMFH procedure for measuring BALnicity was
developed for application to \ion{C}{4} $\lambda$1549 which is not available in
the spectra of the low--redshift quasars.  Application of a modified WMFH
BALnicity procedure to the \ion{Mg}{2} $\lambda$2798 is possible (e.g.  Hall et
al.  2002) but the continuum placement is particularly problematic in this
spectral region where \ion{Fe}{2} emission can vary dramatically from object to
object.  Given a significant number of spectra with both \ion{C}{4}
$\lambda$1549 and \ion{Mg}{2} $\lambda$2798 BAL troughs present, an assessment
of the relation between BALnicity defined using both lines would be possible.
However, given the small number of low--ionization BALs identified from their
\ion{Mg}{2} $\lambda$2798 line properties, none of which possess spectra
including \ion{C}{4} $\lambda$1549, we have chosen not to derive BALnicity
indices for these quasars.

For the majority of candidate BALQSOs, visual inspection coupled, where
appropriate, with the application of the BALnicity determination, produces an
unambiguous classification of an object as a BAL.  Exceptions, as discussed
above, include some quasars with $z\sim 1.2$.  Similarly, given the complex
nature of quasar SEDs in the region of \ion{Mg}{2} it is not always clear
whether apparent absorption features do indicate the presence of broad
absorption as opposed to local minima in the \ion{Fe}{2} emission complexes.  As
a consequence we have assigned a probability ($P({\rm BAL})=$ 1.0, 0.7, 0.3),
essentially equivalent to ``definite'', ``probable'' or ``possible'' BAL
categories, to each quasar identified as a potential BAL.  Table 1 summarises
the properties of the 67 candidate BALQSOs identified in the LBQS.  Column 1
gives the LBQS name, derived from the B1950.0 coordinates; columns 2--3, the
J2000.0 right ascension and declination; column 4, redshift; column 5, $B_J$
magnitude; column 6; probability quasar is a BAL; column 7, BALnicity index,
where available; column 8, classification as a low-- or high--ionization BAL (L
or H) for those objects where spectral coverage of the \ion {Mg}{2}
$\lambda$2798 line is available; column 9, reference to source of the BALnicity
measures.  The BALnicity measures in column 7 have been taken from WMFH and
Korista et al.  (1993) with the average value adopted for objects appearing in
both papers.  The source of the BALnicity measure is denoted in column 9 as ``K''
for Korista et al.  and ``W'' for WMFH.  Additional BALnicity measures come from
application of the WMFH BALnicity estimation procedure to the LBQS spectra.  The
probabilities adopted by WMFH for the ``L'' classifications of two quasars are
included in brackets in column 8.  Comparison of the BALnicity indices for the
objects in common between WMFH and Korista et al.  (1993) indicate that the
uncertainty in the BALnicity values is $\simeq 30\%$.  However, larger errors
can result due to the constraint in the definition of BALnicity that the
absorption be contiguous over $2000\kms$ or more.  Uncertainty in the placement
of the continuum in objects with complex absorption troughs can lead to
discrepant estimates due to the resetting of the integral when the flux rises
above $90\%$ of the continuum level (see WFHM for details).  Extreme examples of
differences in the measured BALnicity values are B1231+1320, with BALnicity\,$=5772\kms$ (WMFH) and BALnicity\,$=1174\kms$
(Korista et al.), and B1138--0126 for which we derive BALnicity\,$=3523\kms$,
whereas Brotherton et al.  (2002) obtain BALnicity\,$=900\pm300\kms$.

\section{THE FRACTION OF BROAD ABSORPTION LINE QUASARS IN THE LBQS}

\subsection{The Observed Fraction of Broad Absorption Line Quasars}

Straightforward estimates of the observed fraction of BALQSOs in the LBQS
can be derived using the sample presented in Table 1 together with the
statistics of the full LBQS. The predicted number of LBQS
quasars within a specified redshift and magnitude range can be calculated 
by summing the factors 

\begin{equation}
N_{quasar} = \sum_{n=1}^{1067} {1\over{z_{comp}}} * {1\over{m_{comp}}} * P
\end{equation}

{\noindent}for all quasars within specified redshift and apparent magnitude
limits.  The redshift completeness factors $z_{comp}$ for each quasar are given
in Table 7 of Hewett et al.  (2001) and the fraction of the LBQS survey area
extending faint enough to include a quasar of a specified magnitude, $m_{comp}$,
is calculated using the magnitude limits and associated areas of sky given in
Table 2 of Hewett et al.  (1995).  To calculate the total number of quasars
within a specified redshift and magnitude range the probabilities, $P$, are set
to unity.  The associated number of BALQSOs is calculated using the
probabilities given for each quasar in column 6 of Table 1.  Table 2 summarizes
the statistics of the fraction of BALQSOs for different redshift and magnitude
ranges.  Section A of the table relates to estimates of the fraction of BALQSOs
calculated from equation (1).  Section B of the table provides information
concerning the intrinsic fraction of BALQSOs, calculated using equation (1) with
the addition of the multiplicative weighting term defined in equation (3) (see
\S 3.2).  Column 1 specifies the redshift range; column 2 specifies the
magnitude range; column 3 lists the total number of quasars satisfying the
redshift and magnitude limits; column 4 gives the number of non--BALQSOs
calculated using equation (1); column 5 specifies the number of BALQSOs used in
the calculations; column 6 gives the number of BALQSOs calculated using equation
(1) (Section A), or, using equation (1) with the addition of the weighting term
defined in equation (3) (Section B); column 7 gives the fraction of BALQSOs,
calculated from columns 4 and 6.

The observed fraction of BALQSOs exhibiting  high--ionization troughs has
been calculated for redshift ranges where \ion{C}{4} $\lambda$1549 is
present in the spectra, i.e. $z \ge 1.3$.  To be absolutely confident
that the presence of the atmospheric cutoff does not affect the
identification of BALs a lower limit of $z=1.5$ has also been employed.
The statistics employing the $z=1.5$ redshift limit are also directly
comparable to a sample of BALQSOs taken from the FIRST survey (Becker et
al. 2000) \---\ see \S 4. An upper redshift limit of $z=3.0$
has been adopted due to the increasingly small fraction of the $B_J$
passband occupied by the quasar SED longward of Lyman $\alpha$
$\lambda$1216 and the corresponding increased uncertainty in the BAL
$k$--correction due to the presence of \ion {N}{5} $\lambda$1240 and
Lyman $\alpha$ $\lambda$1216 absorption troughs. In fact there are only
9 quasars with $z>3.0$ in the LBQS sample, none of which is a BAL, and
their inclusion would make no significant difference to the
statistics.  The straightforward calculation of the observed frequency
yields $0.14 \pm 0.03$ and $0.15\pm 0.03\%$ for lower redshift bounds
of 1.3 and 1.5, respectively (Table 2).

To identify a low--ionization BALQSO from the LBQS discovery spectra it is
necessary that \ion{Mg}{2} $\lambda$2798 is present longward of the
atmospheric cutoff and shortward of $\sim 5600$\AA, where the
signal--to--noise ratio (S/N) of the discovery spectra has declined
significantly from the target of S/N $\simeq 10$ at $4500$\AA. The
redshift range over which the probability of identification of
low--ionization BALQSOs in the LBQS is relatively constant is thus $0.3
\le z \le 1.0$. There are 379 LBQS quasars in this redshift range with
magnitudes $B_J \ge 16.5$. Only 5 quasars are candidate BALQSOs (2 definite;
2 probable; 1 possible), leading to an estimate of the observed frequency of
low--ionization BALQSOs of $\simeq 1\%$ with a large associated error.

Given the relatively small number of BALQSOs in the sample, it is important
to verify that a few objects with large weighting factors
are not contributing disproportionately to the statistics.  In fact,
the redshift completeness factors are small and independent of redshift
within the range $1.3 \le z \le 3.0$ and vary only by $\sim 10\%$ for
$0.3 \le z \le 1.0$. Thus, the deductions concerning the observed
frequency of BALQSOs are virtually independent of any reasonable change to
the redshift completeness factors employed. The magnitude correction
factors are defined accurately by the fraction of the LBQS that extends
to a given magnitude limit. As the entire LBQS extends to $B_J = 18.41$
the area completeness factors are unity brighter than this limit and
the completeness correction remains less than a factor two for all
quasars with $B_J \le 18.64$. However, the LBQS extends fainter than
$B_J=18.77$ in only a few of the 18 fields making up the survey. The
completeness correction is $\simeq 10$ for $18.77 < B_J \le 18.80$ and
reaches $\simeq 20$ for $18.80 < B_J \le 18.85$ where only a single
field reaches $B_J=18.85$. Sixty--five of the 67 candidate BALQSOs in
Table 1 possess magnitude completeness factors $\le 2.6$ but two objects,
1231+1320 and 1240+1607, both of which have magnitudes $B_J = 18.84$,
just brighter than the magnitude limit of the deepest LBQS field, have
completeness factors of $21.6$! Nonetheless, splitting the LBQS into a
bright and faint sample, with the boundary set at $B_J=18.41$,
demonstrates that the inferred fraction of BALQSOs is not unduly
sensitive to the details of the magnitude range employed (see Table
2).

\subsection{The Intrinsic Fraction of Broad Absorption Line Quasars}

To estimate the intrinsic fraction of BALQSOs, it is necessary to
account for the effect of the BAL troughs on the definition of the
flux--limited LBQS sample; effectively the BALQSO $k$--correction.
Specifically, the broad--band magnitudes of BALQSOs appear fainter than
equivalent non--BALQSOs for redshifts where the BAL troughs fall
within the passband defining the broad--band magnitude. For objects
with high values of BALnicity, the size of the effect can be significant.
Quantitative corrections for this effect have been
applied when comparing the properties of BALQSO and non--BALQSOs
(e.g., Stocke et al. 1992).  There is also some evidence that the shapes
of the restframe ultraviolet to optical spectral energy distribution 
of BALQSOs and non--BALQSOs differ, although the effect
may be significant only for the subset of low--ionization BALQSOs
(Sprayberry \& Foltz (1992), Hall et al. (2002), Yamamoto \& 
Vansevi{\v c}ius (1999)).

The $B_J$ passband that defines the LBQS flux--limited sample has an
effective wavelength of $\simeq 4600$\AA \ and extends from $3850$\AA \
to $5300$\AA. The BAL trough associated with \ion {C} {4} $\lambda$1549
enters the passband at $z \simeq 1.5$ and one or more BAL troughs
associated with \ion {C} {4} $\lambda$1549, \ion {Si} {4}
$\lambda$1400, \ion {N} {5} $\lambda$1240 and Lyman $\alpha$
$\lambda$1216 fall within the $B_J$ passband for the entire redshift
range, $1.5 \le z \le 3.0$, used to estimate the frequency of BAL
quasars in \S 2.2. Thus, there is a potentially significant
BALQSO $k$--correction that affects the fraction of BALQSOs
included in the LBQS.

WMFH presented 54 high--quality spectra of BALQSO and non--BALQSOs from the
LBQS, augmented by 17 spectra of BALQSOs from other surveys.  The spectra
have good relative spectrophotometry over a wide wavelength range and so can be
employed to quantify the relative fluxes that would be observed at specified
continuum wavelengths or through particular broad--band filters.  The WMFH
spectra can be employed individually or combined to produce composite spectra.
Two--dimensional BAL $k$--corrections, as a function of redshift and BALnicity,
were constructed using composite BALQSO spectra, as specified in Table 3,
compared to the composite of the WMFH non--BALQSO LBQS quasars.  The selection of
the BALnicity index boundaries was made so that approximately equal numbers of
objects contribute to each composite.  For the most extreme objects,
BALnicity\,$>6000\kms$, it was also possible to generate separate composite
spectra using just the high--ionization and low--ionization objects.  The new
composite spectra were generated using the same procedures as described in WMFH.
Information relating to the composite spectra is summarized in Table 3.  Column
1 is a name, with ``H'' and ``L'' denoting composites constructed from just
high-- and low--ionization BALQSO spectra respectively; column 2 gives the number
of spectra contributing to each composite; column 3 specifies the range of
BALnicity values of the spectra used to generate the composite; column 4 lists
the BALnicity of the composite spectrum.

The non--BALQSO and BALQSO composite spectra are extremely similar at wavelengths
$>1600$\AA.  Normalising the spectra using the average flux over the interval
$1600-2600$\AA \ produces $B_J$ magnitudes all within $0.01\,$mag of each other
at redshift $z=1.3$.  The BAL $k$--corrections are calculated from the
difference in the $B_J$ magnitudes of each BALQSO composite compared to the
non--BALQSO composite at a specified redshift.  The $k$--correction for a
particular BALnicity was derived by interpolating between the loci defined by
the non--BALQSO and the BALQSO composite spectra.  Linear extrapolation was used for
quasars with BALnicity exceeding the BALnicity of the strongest BALQSO composite
used.  The two--dimensional $k$--corrections for the composite spectra are
plotted in Figure 1.  Notwithstanding the substantial range in the appearance of
individual BALQSOs, the $k$--correction calculated from these composite
spectra varies smoothly and is extremely well--behaved, both as a function of
redshift and BALnicity.  It might be supposed that the use of composite spectra
made up of only high--ionization or low--ionization BALQSOs would result in
significantly different $k$--corrections for a specified BALnicity.  Certainly,
the larger the BALnicity index, the greater the probability that an object shows
evidence for low--ionization BALs.  The separate high-- and low--ionization
composite spectra for the objects with BALnicity\,$>6000\kms$ do have very
different mean BALnicity values, $7990\kms$ and $9689\kms$ respectively, but the
results described in this section are completely insensitive to which of the
three BALnicity\,$>6000\kms$ composites (All, High, Low) is used to generate the
$k$--corrections.  In other words, although the loci used to define the
$k$--corrections differ (Figure 1), the interpolated values for a specified
BALnicity and redshift are very similar for objects with BALnicity\,$\la
10000\kms$.  The results of the application of the $k$--corrections calculated
according to this prescription are also not sensitive to the details of the
interpolation.

The reference non--BALQSO spectrum used is the WMFH non--BALQSO composite
spectrum, based on 29 quasars, which is slightly redder than the
composite spectra derived from larger fractions of the LBQS employed in
Hewett et al.  (2001).  However, the much closer match in redshift to
the sample of BALQSOs under consideration make the WMFH--composite
the reference of choice.  Use of the composite from Hewett et al.
(2001) produces small increases in the the inferred intrinsic fractions of BALQSOs
compared to the values given in Table 2.

Making the assumption that the fraction of BALQSOs does not change
significantly close to the LBQS flux limit, an estimate of the true
fraction of BALQSOs that would be included in the survey,
correcting for the depressed apparent magnitudes due to the presence of
the BAL troughs, can be made.  For the 42 LBQS BALQSOs, an additional,
multiplicative, weighting factor

\begin{equation}
W_{c1} = {N_{quasar}(\le 18.85)\over N_{quasar}(\le 18.85-\Delta m_k)}
\end{equation}

{\noindent}is included in equation (1).
$N_{quasar}(\le 18.85)$ is the number of quasars, within some specified
redshift range, to the original LBQS flux limits, $\Delta m_k$ is the
BAL $k$--correction in magnitudes for the quasar and $N_{quasar}(\le
18.85-\Delta m_k)$ is the number of quasars, within the same specified
redshift range, to the revised (brighter) LBQS flux limits for the BAL
quasar. The result of this procedure is an estimated BALQSO fraction of
$19\pm3\%$. To ensure that the calculation is not sensitive to the
exact values of the BAL $k$--corrections, the calculation was repeated
using individual BAL $k$--corrections for 23 of the BALQSOs 
derived directly from the WMFH spectra. Essentially identical results
were obtained.

A somewhat more sophisticated calculation can be performed to allow
for the strong dependence of the BALQSO $k$--correction on redshift.
The importance of allowing for a redshift--dependence can be gauged
by noting that all 5 BALQSOs with BALnicity\,$> 6000\kms$ have
redshifts $1.5 < z < 1.9$, where the effect of the BAL--troughs on
the broad--band $B_J$ magnitudes is smallest. 

Assuming that the fraction of BALQSOs is constant over the redshift interval
$1.5 \le z \le 3.0$, the weighting factor included in equation (1) becomes

\begin{equation}
W_{c2} = {N_{quasar}(\le 18.85)\over N_{quasar}(\le 18.85-\Delta m_k(z))}
\end{equation}

{\noindent}where the quantities are as defined for equation (2), except that
$\Delta m_k(z)$ is now a function of redshift.  Figure 2 illustrates the nature
of the calculation graphically.  The use of the two--dimensional $k$--correction
is effectively the best that can be achieved given the information available.
Applying the weighting factors calculated using equation (3), the estimated
intrinsic BALQSO fraction over the full LBQS magnitude range with $1.5 \le z \le
3.0$ rises to $22\pm4\%$, with one--quarter of the population of BALQSOs
possessing BALnicity\,$> 4000\kms$.  The calculated intrinsic fractions are presented
for three magnitude ranges in Table 2.

The resulting estimate of the intrinsic fraction of BALQSOs is still in
practice a lower limit because it has been assumed that the sample of 42 LBQS
BALQSOs is a fair representation of the parent BALQSO population.  This
is manifestly not the case because the direct relation between BALnicity and the
size of the BAL $k$--correction means that quasars with larger BALnicity values
are increasingly under--represented among the sample of 42 LBQS BALQSOs.  The
procedure described above corrects for the systematic under--representation,
providing that quasars of a given BALnicity appear in the sample.  For the most
extreme objects, those with $10000 < {\rm BALnicity} < 20000\kms$, this is not
the case.  The extreme BALQSO B1232+1325, with redshift $z=2.364$ and
BALnicity\,$=12792\kms$, provides a direct example.  B1232+1325 lies within the
LBQS survey area and possesses an objective--prism spectrum that would make it
one of the most readily detectable quasars in the entire survey.  However, the
broad--band magnitude, $B_J=18.77$, means the quasar falls just below the LBQS
flux--limit in the field.  The BAL $k$--correction, determined directly from the
WMFH--spectrum as described above, is $1.01\,$mag, giving a corrected magnitude
of $B_J=17.76$.  Assuming the fraction of BALQSOs does not vary over the
redshift interval $1.5 \le z \le 3.0$, the effective depth of the LBQS for a
quasar with BALnicity\,$=12792\kms$ is reduced by a factor of 4.5 compared to a
non--BALQSO.  For BALnicity\,$=15000\kms$ the effective depth of the survey is
reduced by a factor of $\simeq 7$, equivalent to only $\simeq 55$ quasars, and
thus the LBQS sample has essentially no sensitivity to BALQSOs with
BALnicity\,$\ge 15000\kms$.

The conclusion that, using the well--defined BALnicity criterion of WMFH, the
proportion of BALQSOs in the redshift range $1.5 < z < 3.0$, with 
corrected magnitudes $16.5 < B_J < 18.75$, is at least one in five ($22\pm4\%$)
is one of the main results of this work. The proportion is essentially
twice the figure generally quoted.

\section{FREQUENCY OF BROAD ABSORPTION LINE QUASARS IN THE FBQS}

The application of the $k$--correction described in \S3.2 leads to a
predicted fraction of BALQSOs of $22\pm4\%$ for samples of quasars
with redshifts $1.5 \le z \le 3.0$ selected according to a continuum
flux--limit at restframe wavelengths $\simeq 2100$\AA.  Assuming that the
intrinsic SEDs of non--BALQSO and BALQSOs are similar over larger
wavelength ranges, then the corrected fraction of BALQSOs in
samples defined at other wavelengths is predicted to be the same. While
there is clear evidence at X--ray wavelengths for the presence of
additional absorbing material (Green et al.  1995, 2001), more
observations of larger samples are required before a meaningful
comparison of the fraction of BALQSOs determined at restframe
X--ray and ultraviolet wavelengths is possible.  At the opposite
wavelength extreme, the lack of BALQSOs detected as strong
radio--sources led to suggestions that the SEDs of non--BALQSO and BALQSOs
at radio wavelengths were intrinsically different (Stocke et
al. 1992). The existence of this potentially important difference in
the SEDs of the two classes of quasars has been challenged by Becker et
al. (2001) who discuss the incidence of BALQSOs in the FBQS survey
(White et al. 2000), concluding that ``The frequency of BALQSOs in
the FBQS is significantly greater, perhaps by as much as a factor of 2,
than that inferred from optically--selected samples.''

The conclusion of Becker et al. relies on the adoption of a BALQSO
fraction for optically--selected samples of only $10\%$, combined with
an extended definition of ``BAL'' quasars in the FBQS that differs from
that applied in the optical samples.  Employing a consistent definition
for the identification of BALQSOs in both samples and incorporating the
revised estimate for the incidence of BALQSOs from \S3.2 leads to a very
different conclusion.

Of the 29 quasars presented in Becker et al., 17 have redshifts $1.5 \le z \le
3.0$ and magnitudes $16.0 \le E \le 17.8$.  Four of these objects possess a
BALnicity index of zero and should not be included when comparing to samples
which, by definition, require BALnicity\,$>0$.  Therefore, Becker et al.'s sample
contains 13 objects with positive BALnicity that could have been identified from
the LBQS discovery spectra.  There are 142 quasars, of all types, in the FBQS
within the same redshift and magnitude range, leading to an estimate of the
fraction of BALQSOs of $9\pm3\%$.  In contrast to the $B_J$ passband, that
defines the LBQS selection, the Palomar $E$ passband ($\lambda_{eff} \simeq
6400$\AA, $\Delta\lambda \simeq 400$\AA) samples the same restframe wavelengths
($1600-2600$\AA) at redshifts $z \sim 2$ used to match the SEDs of non--BALQSO 
and
BALQSO SEDs in \S 3.  The $E$ passband is also unaffected by the presence of both
high--ionization and low--ionization BAL troughs for redshifts $1.5 \le z \le
3.0$, so the BAL $k$--correction is essentially zero and the observed fraction
is equivalent to the intrinsic fraction of BALQSOs.

As a consistency check, it is possible to construct directly a Palomar
$E$--band--limited quasar sample, an optically--selected equivalent to the FBQS,
using the LBQS sample.  The procedure is described in detail in Hewett et al.
(2001) but an outline is included here.  Broadband colours were calculated using
the Palomar $O$ and $E$ sensitivity curves of Minkowski \& Abell (1963) and our own
determination of the $B_J$ sensitivity.  The filter plus emulsion sensitivity
curves were combined with one reflection off aluminum and atmospheric absorption
and extinction appropriate for observations made at an airmass of 1.3 for a
relatively low--altitude site such as Siding Spring or Mount Palomar.  Synthetic
photometry was performed using the {\tt synphot} package in the Space Telescope
Science Data Analysis System.

The measured $B_J-E$ and $O-E$ colours for the WMFH non--BALQSO composite spectrum,
as a function of redshift, can be used to estimate an $E$--magnitude and $O-E$
colour for each non-BALQSO in the LBQS.  For the BALQSOs a directly analogous
procedure to that employed to produce the two--dimensional $k$--correction in
\S3.2 is used.  The $B_J-E$ and $O-E$ colours for the BALQSO composite spectra
are calculated as a function of redshift.  The resulting loci in the $B_J-E$ and
$O-E$ {\it versus} $z$ planes are then used to provide an $E$--magnitude and
$O-E$ colour for each of the 42 BALQSOs, with given BALnicity and $z$, via
interpolation.  The pseudo--$E$--limited sample is then derived by requiring the
quasars to have magnitudes in the range $16.0 \le E \le 17.8$ and $O-E \le 2.0$,
i.e., the magnitude selection limits of the FBQS, excluding the very brightest
objects.  A limit of $E < 16.0$ is used because the majority of quasars brighter
than this limit would not satisfy the $B_J \le 16.5$ limit of the LBQS.  The
result of the procedure applied to the 375 LBQS quasars is a sample of 225
quasars, of which 45 are BALQSOs, giving a predicted BALQSO fraction of $20\pm3\%$.
The result is insensitive to which of the BALnicity\,$>6000\kms$ composite
spectra are used because whether the few objects with large BALnicity fall
within the $E$ flux limits does not change as different composite spectra are
used to calculate the $B_J-E$ colours.  The proportion of BALQSOs predicted for
an optically-selected sample equivalent to the FBQS derived using this direct
procedure is expected to be very similar to the $22\%$ derived in \S3.2 because
the $E$--band flux limit is equivalent to the $1600-2600$\AA \ restframe region
used to normalise the SEDs of the non--BALQSO and BALQSO spectra.

The results of the comparison do not apply to BALQSOs with extreme
BALnicities, BALnicity\,$\ga 15000\kms$, because of the lack of sensitivity of the
LBQS to such BALQSOs.  The $E$--band flux limit that defines the FBQS is not
so affected but the requirement that objects have $O-E \le 2.0$ will result in
the exclusion of BALQSOs with BALnicity\,$\ga 15000\kms$ at redshifts $\ga 2.6$.
However, the results of a like for like comparison between the fraction of BAL
quasars in the LBQS and the FBQS is that the BALQSOs are only half as common
in the FBQS.  While larger samples of objects, from SDSS, will improve the
accuracy of this figure, the result is statistically significant and
produces a very different interpretation to that of Becker et al.

\section{DISCUSSION}

\subsection{The Fraction of BALQSOs in the LBQS}

A primary conclusion of this paper is that over the redshift range $1.5 \le z
\le 3.0$ the observed fraction of BALQSOs in the LBQS is $15\pm3\%$.  This
result involves no correction for differences in the SEDs between non--BALQSO and
BALQSOs.  Application of a well--determined BAL $k$--correction to allow for the
differences in the SED shortward of $\sim 2000$\AA \ results in an estimate of
the intrinsic fraction of BALQSOs of $22\pm4\%$.  These results supersede those
given in Foltz et al.  (1990) which were based on a preliminary version of the
LBQS catalogue.  The higher fraction of BALQSOs, both observed and inferred, in a
bright optically--selected sample has significant implications for the
interpretation of recent results concerning the fraction of BALQSOs in the
FBQS (Becker et al.  2001) and the SDSS EDR (Tolea et al.  2002) since both papers
use an estimate of $\simeq 10\%$ for the fraction of BALQSOs in the LBQS.

\subsection{Redshift Dependence of the BALQSO Fraction}

There are insufficient objects in the LBQS sample to place useful constraints on
any redshift dependence of the BALQSO fraction.  However, the distribution of
objects with redshift is consistent with no $z-$dependence.  This contrasts with
the results of Tolea et al., who present evidence in their Figure 1 for a
significant variation in the BALQSO fraction with redshift.  Tolea et al.  note
specifically that selection effects in the construction of the SDSS EDR sample
may be important.  It is noteworthy that the systematic rise in the BAL
fraction, from $\sim 10\%$ at $z=2.0$ to a peak of $\sim 30\%$ at $z=2.7$,
followed by a decline to $\sim 15\%$ at $z=3.2$, coincides with the strong
change in the effectiveness of the SDSS quasar colour--selection algorithm
(Richards et al.  2002; Figures 10 \& 15).  If the presence of BAL troughs moves
quasars further (closer) from the stellar locus in colour--space then the
apparent BALQSO fraction will rise (fall).  Detailed simulations to quantify the
effectiveness of the final quasar--selection algorithm as a function of redshift
and quasar SED will no doubt be undertaken as part of the SDSS.  We conclude
that there is no strong evidence favouring a significant change in the fraction
of BALQSOs over the redshift range $1.5 \la z \la 3.5$ but that forthcoming
investigations utilising the quasar catalogue from the SDSS will resolve the
issue.

\subsection{The BALQSO Fraction in the SDSS EDR and FBQS}

The inferred incidence of BALQSOs from the LBQS, $22\pm4\%$, is somewhat
higher than the average value of $15\pm1\%$ from Tolea et al.  Given the smaller
number of objects in the LBQS sample and the unquantified selection effects in
the SDSS EDR sample, a figure of $\simeq 20\%$ for the intrinsic fraction of BAL
quasars in a flux limited sample at a restframe wavelength of $\sim 2100$\AA \
over the redshift range $z \sim 1.5-3.5$ is consistent with both surveys.  The
doubling of the inferred fraction of BALQSOs also reduces the apparent
significance of the anomalously high fraction (7 out of 20 quasars) of BALQSOs
among gravitational lenses to which Chartas (2000) draws attention.

The calculation of the intrinsic fraction of BALQSOs in both investigations is
predicated on the restframe wavelengths used to define the flux--limited samples
of non--BALQSO and BALQSOs.  In the case of the LBQS investigation this
restframe wavelength is $\simeq 2100$\AA.  For the Tolea et al.  investigation
the wavelength is very similar, $\sim 2000$\AA \ (from the effective wavelength
of the SDSS $i$--band and the typical redshift of the Tolea et al.  quasars).
If, in fact the SEDs of BALQSOs are redder than their non--BAL counterparts
longward of $\sim 2100$\AA \, then the intrinsic fraction of BALQSOs increases.
It should be stressed that the two--dimensional $k$--correction employed in
\S3.2 incorporates any differences in the SEDs of non--BALQSO and BALQSOs
shortward of $\sim 2100$\AA \ but that it is extremely unlikely that even the
continuum SEDs are identical at longer wavelengths.  Sprayberry \& Foltz (1992)
and Yamamoto \& Vansevi{\v c}ius (1999) both find that the differences between
the SEDs of luminous non--BALQSO and BALQSOs selected in optical surveys can be
explained by the presence of relatively small amounts of reddening by dust with
properties similar to that in the Milky Way and the Magellanic Clouds.  If such
modest levels of reddening are representative of the BALQSO population as a
whole then the differences in SEDs at longer wavelengths are relatively small.
On the other hand, there are individual examples of reddened non--BALQSO and
BALQSOs in the SDSS (e.g.  Gregg et al.  2002; Hall et al.  2002) and even in
the LBQS (B0059--2735; see \S 5.4).  Compared to the LBQS, the SDSS, with its
object selection, based on the $i$--band, at slightly longer restframe
wavelength, combined with increased depth (the $i\le 19.1$ limit reaches
$\sim0.5\,$mag fainter down the luminosity function for typical quasars with
redshifts $z\simeq 2-3$) is more sensitive to quasars experiencing small amounts
of reddening.  However, both the LBQS and SDSS are largely insensitive to
quasars with even moderate amounts of reddening.  For example, the presence of
dust producing an $E(B-V)$ of only $0.25\,$mag, acting as a screen close to the
quasar, results in a bright, $i=17.0$, redshift $z=2.5$, quasar falling below
the SDSS $i$--band flux limit.  Quantifying the space density of objects
suffering even modest amounts of reddening will likely require a wide--field
quasar sample defined at near--infrared wavelengths.

One can ask whether the numbers of BALQSOs found in the LBQS, that are also
detected in the FIRST survey, are consistent with the proportion of BALQSOs
found in the FBQS.  Table 4 provides summary information for the 5 LBQS BAL
quasars with counterparts in the FIRST catalogue.  Column 1 is the LBQS name;
column 2, redshift; column 3, $B_J$ magnitude; column 4, probability the object
is a BALQSO; column 5, separation, in arcseconds, between the optical
position and the FIRST radio source; column 6, the integrated flux taken from
the FIRST catalogue.  The information in columns 1--4 is identical to the
corresponding entries in Table 1.

Hewett et al.  (2001) showed that the fraction of LBQS
quasars included in the FIRST catalogue is $12\%$.  Of the 42 candidate BAL
quasars, with $1.5 \le z \le 3.0$, 28 lie within the boundaries of the currently
available FIRST survey.  If the probability an LBQS quasar is detected as a
FIRST source is independent of the BAL properties then the expected number of
BALQSOs detected as FIRST sources is 3.4.  From the comparison presented in
\S3.2 the probability a BALQSO is also a FIRST source is somewhat under half
that for a non--BALQSO, leading to a predicted number of BALQSOs
detected as FIRST sources of $\la 2$.  In fact, one quasar, the low--ionization
object B1331--0108 (Table 4), is detected in the FIRST survey.  The numbers are
extremely small, illustrating the difficulty of establishing the distribution of
radio--properties of BALQSOs, even with quite substantial sample sizes, but
are consistent with the conclusions of \S3.2, that BALQSOs are approximately
half as likely to be detected as FIRST sources compared to their non--BALQSO
counterparts.

Small numbers also limit what can be concluded from Tolea et al.'s analysis of
their SDSS sample but their relative fraction, $5/116 \simeq 5\%$, of BALQSOs
detected as FIRST sources is in good agreement with our results.  In their
discussion of the SDSS BALQSOs that are also FIRST sources, Menou et al.  (2001)
have already pointed out that the observed fraction of radio--detected BALQSOs
is inconsistent with the claim of Becker et al.  (2000) that BALQSOs are more
common in the FBQS than in optically--selected samples.  Our results, together
with those of Tolea et al., based on samples defined using the BALnicity
criterion, strengthen this conclusion, producing a consistent picture in which
BALQSOs are approximately half as likely as non--BALQSOs to possess detections
in FIRST.

A further important consistency check, to verify that the relation
between the BALQSOs with $z> 1.5$ in the LBQS and FBQS is well
understood, would be to verify that the SEDs of the BALQSOs in the
FBQS are not significantly different from the optically--selected
objects used to derive the two--dimensional $k$--correction of \S3.2.
Unfortunately, the relatively long wavelength ($\simeq 4000$\AA) of the
blue edge of the FBQS spectra, combined with the low redshift of many
of the $z> 1.5$ BALQSOs means there is insufficient restframe wavelength
coverage to perform a useful comparison.

\subsection{LoBALs and FeLoBALs}

Following the identification of the prototype object, B0059--2735 (Hazard et al.
1987), a key finding of the FBQS survey (Becker et al.  1997) and more recently
of the SDSS survey (Hall et al.  2002) has been the definition of a subset of
the low--ionization BALQSOs (``LoBALs'') that show strong and complex absorption
from low--ionization species including excited states of Fe+.  Extreme examples
of these objects, often referred to as FeLoBALs are given by Hall et al.
(2002).  Notwithstanding the inclusion of the prototype FeLoBAL, B0059-2735, in
the LBQS, optical surveys with relatively bright flux--limits at blue
wavelengths are largely insensitive to the detection of such objects because of
their very large BAL $k$--corrections in blue passbands.

Given the tiny number of objects, analysis of the statistics of BALQSOs in
the redshift interval, $0.3 \le z \le 1.0$, where \ion{Mg}{2} $\lambda2798$\AA \
broad absorption is detectable, has been confined to the calculation of the
observed fraction in the LBQS (\S 3.1).  There is little more quantitative that
can be said, although, of the 4 LBQS candidate BALQSOs with $0.3 \le z \le
1.0$ within the FIRST survey boundaries, 2 are detected as FIRST sources,
B1016--0248 and B1235+1807B (Table 4).  Extending the redshift range, to include
all 25 LBQS candidate BALQSOs not in the redshift range $1.5 \le z \le 3.0$,
produces 4 detections out of the 21 objects within the FIRST survey boundaries.
The 4 quasars are B1016--0248, B1235+1807B, B0059---0205 and B1138--0126 (Table
4).  The quasar B1138--0126, also a low--ionization BALQSO, is in fact a
Fanaroff--Riley class II, radio--loud source, that has been discussed by
Brotherton et al.  (2002).  Thus, 4 of 21 objects are detected in FIRST and the
3 objects with reliable ionization classifications are all low--ionization BALQSOs.
The statistics suggest that there may be an affinity between the presence of
low--ionization BALQSOs and radio emission at intermediate radio power.  However, much
larger samples are required to establish such a relationship.

\subsection{A Radio/BALQSO Dichotomy?}

One very important result of the FBQS has been to question and indeed break the
apparent dichotomy between the presence of BAL troughs and strong radio
emission.  In this paper the phrase ``FIRST detection'' has been used instead of
``radio loud'' since the former does not necessarily imply the latter.  Clearly,
samples of objects satisfying any classical definition of ``radio loud'' do
contain unambiguous examples of BALQSOs (e.g., Gregg et al.  2000;
Brotherton et al 2002).  However, the statistics of samples defined according to
positive values of BALnicity in the redshift range $1.5 \la z \la 3.5$ show that
the frequency of BALQSOs with $S_{1.4{\rm GHz}}\ga 1\,$mJy is approximately
half that for non--BALQSOs.  This conclusion is consistent with previous
studies of the incidence of BALQSOs detectable at radio wavelengths (e.g.,
Brotherton et al.  1998), given the very small number of detections involved.

As noted above, the issue of the ``radio avoidance'' of BALQSOs is
not a simple binary question, i.e., it is clear that some bona fide BAL
quasars are radio loud.  Furthermore, evidence is also mounting that
the ``associated absorbers'' often seen in the rest ultraviolet spectra
of radio loud quasars may be signaling the presence of outflows that
are related to those of BAL flows (see papers in Crenshaw et al.
2002).  As noted by Weymann (2002), it is no longer profitable to argue
over whether a specific absorption system is truly a BAL trough or
whether a specific BALQSO is radio loud. Instead, efforts should be
concentrated on understanding the connection between radio power and
the character of the outflows.  One strategy for this line of research
would be to examine the ultraviolet absorption properties of very
powerful radio quasars.  To that end, we are currently analyzing a
moderately large sample of extremely radio loud quasars in hopes of
shedding some light on this issue (Weymann et al., in preparation).

We thank our friend and longtime collaborator, Ray Weymann, for his frequent
words of encouragement.  We are grateful to Xiaohui Fan for providing digital
versions of the SDSS passbands.  We are grateful to an anonymous referee for
a careful reading of the manuscript. This research has made use of the NASA/IPAC
Extragalactic Database (NED) which is operated by the Jet Propulsion Laboratory,
California Institute of Technology, under contract with the National Aeronautics
and Space Administration.  We are pleased to acknowledge the continued support
provided for the LBQS through NSF grant AST 98--03072.  Data and analysis
facilities at the Institute of Astronomy were provided in part by the Starlink
Project which is run by CCLRC on behalf of PPARC.

\clearpage

\begin{deluxetable}{lcccccrcc}
\tabletypesize{\footnotesize}
\tablewidth{0pt}
\tablenum{1}
\tablecaption{LBQS Broad Absorption Line Quasars}

\tablehead{
\colhead{LBQS Name}  & 
\colhead{R.A.} & 
\colhead{Dec.} & 
\colhead{$z$} &
\colhead{$B_J$} &
\colhead{Probability} &
\colhead{BALnicity} & 
\colhead{Class} &
\colhead{Reference}
\\
\colhead{} & 
\colhead{(J2000.0)} & 
\colhead{(J2000.0)} & 
\colhead{} & 
\colhead{} & 
\colhead{$P({\rm BAL})$} &
\colhead{($\kms$)} &
\colhead{} &
\colhead{}
}
\startdata
B0004+0147 & 00 07 22.50 & +02 04 12.5 & 1.710 & 18.13 & 1.0 & 255 & \nodata & K \\
B0009+0219 & 00 12 19.67 & +02 36 35.7 & 2.642 & 17.99 & 1.0 & 1738 & \nodata & \nodata \\
B0018+0047 & 00 21 27.91 & +01 04 19.9 & 1.835 & 17.82 & 1.0 & 329 & \nodata & K \\
B0019+0107 & 00 22 27.48 & +01 24 12.7 & 2.130 & 18.09 & 1.0 & 2305 & H & K,W \\
B0020-0154 & 00 23 02.34 & --01 38 16.2 & 1.460 & 18.27 & 1.0 & 1502 & \nodata & \nodata \\
B0021-0213 & 00 24 10.90 & --01 56 47.1 & 2.348 & 18.68 & 1.0 & 5179 & H & K,W \\
B0022+0150 & 00 24 35.35 & +02 06 48.3 & 2.826 & 18.35 & 1.0 & 224 & \nodata & K \\
B0025-0151 & 00 27 33.80 & --01 34 52.3 & 2.076 & 18.06 & 1.0 & 2878 & H & K,W \\
B0029+0017 & 00 31 35.58 & +00 34 21.1 & 2.253 & 18.64 & 1.0 & 5263 & H & K,W \\
B0045-2606 & 00 48 12.56 & --25 50 04.2 & 1.242 & 18.05 & 0.7 & 292 & \nodata & \nodata \\
B0049-0123 & 00 51 35.29 & --01 07 09.4 & 1.560 & 17.84 & 0.7 & 28 & \nodata & \nodata \\
B0049-2535 & 00 52 11.07 & --25 18 58.3 & 1.528 & 18.52 & 0.3 & 1936 & \nodata & \nodata \\
B0051-0019 & 00 53 55.14 & --00 03 09.4 & 1.713 & 18.67 & 1.0 & 3244 & \nodata & K \\
B0054+0200 & 00 56 44.65 & +02 16 30.1 & 1.872 & 18.65 & 1.0 & 498 & \nodata & K \\
B0059-0206 & 01 02 05.60 & --01 50 38.5 & 1.321 & 18.76 & 1.0 & \nodata & \nodata & \nodata \\
B0059-2735 & 01 02 17.04 & --27 19 50.0 & 1.593 & 18.13 & 1.0 & 11053 & L & K,W \\
B0103-2753 & 01 05 34.75 & --27 36 58.2 & 0.848 & 18.07 & 1.0 & \nodata & L & \nodata \\
B0106-0113 & 01 08 55.03 & --00 57 47.1 & 1.668 & 18.07 & 1.0 & 377 & \nodata & \nodata \\
B0109-0128 & 01 12 27.59 & --01 12 21.8 & 1.758 & 18.32 & 1.0 & 399 & \nodata & \nodata \\
B1009+0222 & 10 11 49.00 & +02 07 31.8 & 1.349 & 18.62 & 1.0 & 1565 & \nodata & K \\
B1016-0248 & 10 19 00.86 & --03 03 50.3 & 0.717 & 18.46 & 0.3 & \nodata & L & \nodata \\
B1029-0125 & 10 31 49.51 & --01 41 11.1 & 2.029 & 18.68 & 1.0 & 1848 & H & K,W \\
B1133+0214 & 11 36 31.88 & +01 58 00.6 & 1.468 & 18.38 & 1.0 & 1950 & \nodata & \nodata \\
B1136-0109 & 11 39 04.42 & --01 26 24.8 & 1.378 & 18.48 & 0.3 & 189 & \nodata & \nodata \\
B1138-0126 & 11 41 11.62 & --01 43 06.6 & 1.266 & 18.52 & 1.0 & 3523 & L & \nodata \\
B1203+1530 & 12 06 26.14 & +15 13 35.4 & 1.628 & 18.70 & 1.0 & 1517 & \nodata & \nodata \\
B1203+1703 & 12 06 20.17 & +16 46 39.1 & 1.401 & 18.71 & 0.7 & 1052 & \nodata & \nodata \\
B1205+1436 & 12 08 25.34 & +14 19 20.9 & 1.643 & 18.38 & 1.0 & 788 & H & K,W \\
B1208+1535 & 12 11 25.46 & +15 18 51.5 & 1.961 & 17.93 & 1.0 & 4545 & H & K,W \\
B1212+1445 & 12 14 40.28 & +14 28 59.5 & 1.627 & 17.87 & 1.0 & 3618 & H & K,W \\
B1214+1753 & 12 16 56.88 & +17 37 13.3 & 0.679 & 17.65 & 0.7 & \nodata & L & \nodata \\
B1216+1103 & 12 19 30.93 & +10 47 00.9 & 1.620 & 18.28 & 1.0 & 4791 & H & K,W \\
B1219+1244 & 12 22 21.75 & +12 28 20.0 & 1.309 & 18.66 & 1.0 & 3138 & \nodata & \nodata \\
B1224+1349 & 12 26 35.58 & +13 32 51.6 & 1.838 & 18.18 & 1.0 & 420 & \nodata & \nodata \\
B1228+1216 & 12 31 16.43 & +12 00 24.1 & 1.408 & 17.54 & 1.0 & 496 & \nodata & K \\
B1230+1705 & 12 33 10.69 & +16 49 05.8 & 1.420 & 18.44 & 1.0 & 2945 & \nodata & \nodata \\
B1231+1320 & 12 33 55.62 & +13 04 08.9 & 2.380 & 16.86 & 1.0 & 3473 & L(0.67) & K,W \\
B1234+0122 & 12 37 24.55 & +01 06 15.2 & 2.025 & 18.00 & 1.0 &    4 & H & W \\
B1235+0216 & 12 38 13.01 & +02 00 19.6 & 0.672 & 17.65 & 0.7 & \nodata & L & \nodata \\
B1235+0857 & 12 37 54.82 & +08 41 06.4 & 2.898 & 18.17 & 1.0 & 815 & H & K,W \\
B1235+1453 & 12 37 36.40 & +14 36 40.5 & 2.699 & 18.56 & 1.0 & 2657 & H & K,W \\
B1235+1807B & 12 38 20.22 & +17 50 38.0 & 0.449 & 16.86 & 1.0 & \nodata & L & \nodata \\
B1236+0128 & 12 39 11.52 & +01 12 13.6 & 1.262 & 17.64 & 0.3 & \nodata & \nodata & \nodata \\
B1239+0028 & 12 42 02.65 & +00 12 28.8 & 1.214 & 17.46 & 1.0 & 1733 & \nodata & \nodata \\
B1239+0955 & 12 41 35.89 & +09 39 31.5 & 2.013 & 18.38 & 1.0 & 708 & \nodata & K \\
B1239-0231 & 12 41 57.33 & --02 47 32.1 & 1.234 & 17.72 & 0.3 & \nodata & \nodata & \nodata \\
B1240+1607 & 12 43 03.62 & +15 50 47.6 & 2.360 & 18.84 & 1.0 & 2867 & H & K,W \\
B1243+0121 & 12 45 51.45 & +01 05 05.0 & 2.796 & 18.50 & 1.0 & 5953 & H & K,W \\
B1314+0116 & 13 17 14.23 & +01 00 13.0 & 2.686 & 18.65 & 1.0 & 2626 & H & K,W \\
B1326-0249 & 13 28 50.00 & --03 04 50.4 & 1.407 & 18.74 & 0.3 & \nodata & \nodata & \nodata \\
B1331-0108 & 13 34 28.05 & --01 23 48.8 & 1.881 & 17.87 & 1.0 & 7911 & L & K,W \\
B1429-0036 & 14 31 43.77 & --00 50 11.9 & 1.180 & 17.76 & 1.0 & 1500 & \nodata & \nodata \\
B1442-0011 & 14 45 14.84 & --00 23 58.4 & 2.226 & 18.24 & 1.0 & 5142 & H & K,W \\
B1443+0141 & 14 45 45.29 & +01 29 12.4 & 2.451 & 18.20 & 1.0 & 7967 & H & K,W \\
B2111-4335 & 21 15 06.97 & --43 23 09.5 & 1.708 & 16.68 & 1.0 & 7249 & \nodata & \nodata \\
B2116-4439 & 21 20 11.66 & --44 26 53.8 & 1.480 & 17.68 & 1.0 & 2594 & \nodata & \nodata \\
B2140-4552 & 21 43 28.91 & --45 38 50.7 & 1.688 & 18.30 & 1.0 & 1410 & \nodata & \nodata \\
B2154-2005 & 21 57 05.92 & --19 51 13.6 & 2.035 & 18.12 & 1.0 & 962 & H & K,W \\
B2201-1834 & 22 04 01.61 & --18 19 42.0 & 1.814 & 17.81 & 1.0 & 1612 & H & K,W \\
B2208-1720 & 22 11 15.47 & --17 05 25.4 & 1.210 & 17.65 & 1.0 & 4271 & \nodata & \nodata \\
B2210-1751 & 22 13 09.65 & --17 37 01.5 & 1.557 & 17.86 & 0.3 & \nodata & \nodata & \nodata \\
B2211-1915 & 22 14 37.90 & --19 00 57.3 & 1.952 & 18.02 & 1.0 & 27 & \nodata & W \\
B2212-1759 & 22 15 31.68 & --17 44 08.7 & 2.217 & 17.94 & 1.0 & 2221 & \nodata & K \\
B2239+0007 & 22 41 47.34 & +00 22 54.4 & 1.440 & 18.28 & 0.3 & \nodata & \nodata & \nodata \\
B2241+0016 & 22 44 31.49 & +00 32 25.8 & 1.394 & 18.30 & 1.0 & 396 & \nodata & \nodata \\
B2350-0045A & 23 52 53.50 & --00 28 50.7 & 1.617 & 18.63 & 1.0 & 6964 & L(0.27) & K,W \\
B2358+0216 & 00 01 21.70 & +02 33 04.9 & 1.872 & 18.61 & 1.0 & 6283 & \nodata & K \\
\enddata
\end{deluxetable}

\begin{deluxetable}{ccccccc}
\tabletypesize{\footnotesize}
\tablewidth{0pt}
\tablenum{2}
\tablecaption{Observed and Intrinsic Fractions of Broad Absorption Line Quasars in the LBQS}

\tablehead{
\colhead{Redshift Interval}  & 
\colhead{Magnitude Interval} & 
\colhead{$n_{total}$} & 
\colhead{$\sum$ non--BAL} &
\colhead{$n_{BAL}$} &
\colhead{$\sum$ BAL} &
\colhead{Fraction}
}
\startdata
\multicolumn{7}{c}{A. Observed Fractions} \\
& & & & & & \\
1.5 - 3.0 & 16.50 - 18.85 & 375 & 537.9 & 40.3 & 97.1  & 0.15$\pm0.03$ \\
1.3 - 3.0 & 16.50 - 18.85 & 475 & 687.5 & 50.9 & 112.0 & 0.14$\pm0.03$ \\
& & & & & & \\
1.5 - 3.0 & 16.50 - 18.41 & 221 & 209.8 & 28.0 & 30.4  & 0.13$\pm0.03$ \\
1.3 - 3.0 & 16.50 - 18.41 & 281 & 268.1 & 34.3 & 37.3  & 0.12$\pm0.03$ \\
& & & & & & \\
1.5 - 3.0 & 18.42 - 18.85 & 154 & 328.1 & 12.3 & 66.4  & 0.17$\pm0.04$ \\
1.3 - 3.0 & 18.42 - 18.85 & 194 & 419.4 & 16.6 & 74.8  & 0.15$\pm0.04$ \\
& & & & & & \\
0.3 - 1.0 & 16.50 - 18.85 & 379 & 643.2 & 3.6 & 5.8 & 0.01\\
& & & & & & \\
\multicolumn{7}{c}{B. Intrinsic Fractions} \\
& & & & & & \\
1.5 - 3.0 & 16.50 - 18.85 & 375 & 537.9 & 40.3 & 152.8 & 0.22$\pm0.04$ \\
1.5 - 3.0 & 16.50 - 18.41 & 221 & 209.8 & 28.0 & 46.2  & 0.18$\pm0.03$ \\
1.5 - 3.0 & 18.42 - 18.85 & 154 & 328.1 & 12.3 & 106.6 & 0.24$\pm0.04$ \\
\enddata
\end{deluxetable}

\begin{deluxetable}{cccc}
\tabletypesize{\footnotesize}
\tablewidth{0pt}
\tablenum{3}
\tablecaption{Composite Spectra from the WMFH Quasar Sample}

\tablehead{
\colhead{Name}  & 
\colhead{$n_{quasar}$} & 
\colhead{BALnicity Interval} & 
\colhead{Mean BALnicity}
\\
\colhead{} & 
\colhead{} & 
\colhead{$(\kms)$} & 
\colhead{$(\kms)$} 
}
\startdata
Com\_non--BAL & 29 & 0 & 0\\
Com\_BAL\_1   & 9   & 1501-3000 & 2337\\
Com\_BAL\_2   & 15   & 3001-6000 & 5093\\
Com\_BAL\_3  & 11 &  $\ge$6001 & 8669\\
Com\_BAL\_3\_H &  6.5 &  $\ge$6001 & 7990\\
Com\_BAL\_3\_L &  4.5 &  $\ge$6001 & 9689\\
\enddata
\end{deluxetable}

\begin{deluxetable}{cccccc}
\tabletypesize{\footnotesize}
\tablewidth{0pt}
\tablenum{4}
\tablecaption{LBQS BALQSO Candidates with FIRST Detections}

\tablehead{
\colhead{LBQS Name}  & 
\colhead{$z$} & 
\colhead{$B_J$}  & 
\colhead{$P({\rm BAL})$} & 
\colhead{Separation} &
\colhead{$S_{1.4{\rm GHz}}$}
\\
\colhead{} & 
\colhead{} & 
\colhead{} & 
\colhead{} &
\colhead{(arcsec)} &
\colhead{(mJy)}
}
\startdata
B0059--0205 & 1.321 & 18.76 &  1.0 & 0.4 & $1.81\pm0.15$ \\
B1016--0248 & 0.717 & 18.46 &  0.3 & 0.2 & $3.29\pm0.16$ \\
B1138--0216 & 1.266 & 18.52 &  1.0 & 8.1\tablenotemark{a} & $252\pm0.5$\\
B1235+1807B & 0.449 & 16.86 &  1.0 & 1.0 & $6.07\pm0.14$ \\
B1331--0108 & 1.881 & 17.87 &  1.0 & 0.3 & $2.99\pm0.14$ \\
\enddata
\tablenotetext{a}{Distance to closest of the two associated radio--lobes.}
\end{deluxetable}
 
\clearpage

\center{\bf FIGURE CAPTIONS}

\figcaption{BAL $k$--correction, in magnitudes, as a function of redshift.
The individual data points were derived from the difference in $B_J$
magnitude for each of the 5 composite BALQSO spectra, listed in Table 3,
compared to the composite non--BALQSO spectrum. The individual data points
have been joined to guide the eye. The locus corresponding to a
particular composite spectrum can be identified using the values of
the mean BALnicity given in column 4 of Table 3.} 

\figcaption{Graphical illustration of the procedure used to calculate
the fraction of BALQSOs in the sample after correcting for the 
$k$--correction due to the difference in SEDs shortward of $2100$\AA. 
The distribution of quasars
($\bullet$) within specified redshift and magnitude limits (dashed
rectangle) is shown. The open symbol denotes a BALQSO, the
$k$--corrected magnitude of which is indicated by the top of the upper
arrow. The dot--dash line shows the $k$--corrected magnitude for the
quasar at all redshifts. The corrected limiting magnitude for non--BALQSOs 
(solid line) that corresponds to the detection of the BALQSO 
at the faint limit of the sample also moves brightward by the same
amount (lower arrow). The ratio of the total number of non--BALQSOs 
to the number of non--BALQSOs within the shaded region determines
the weighting applied to the BALQSO.}

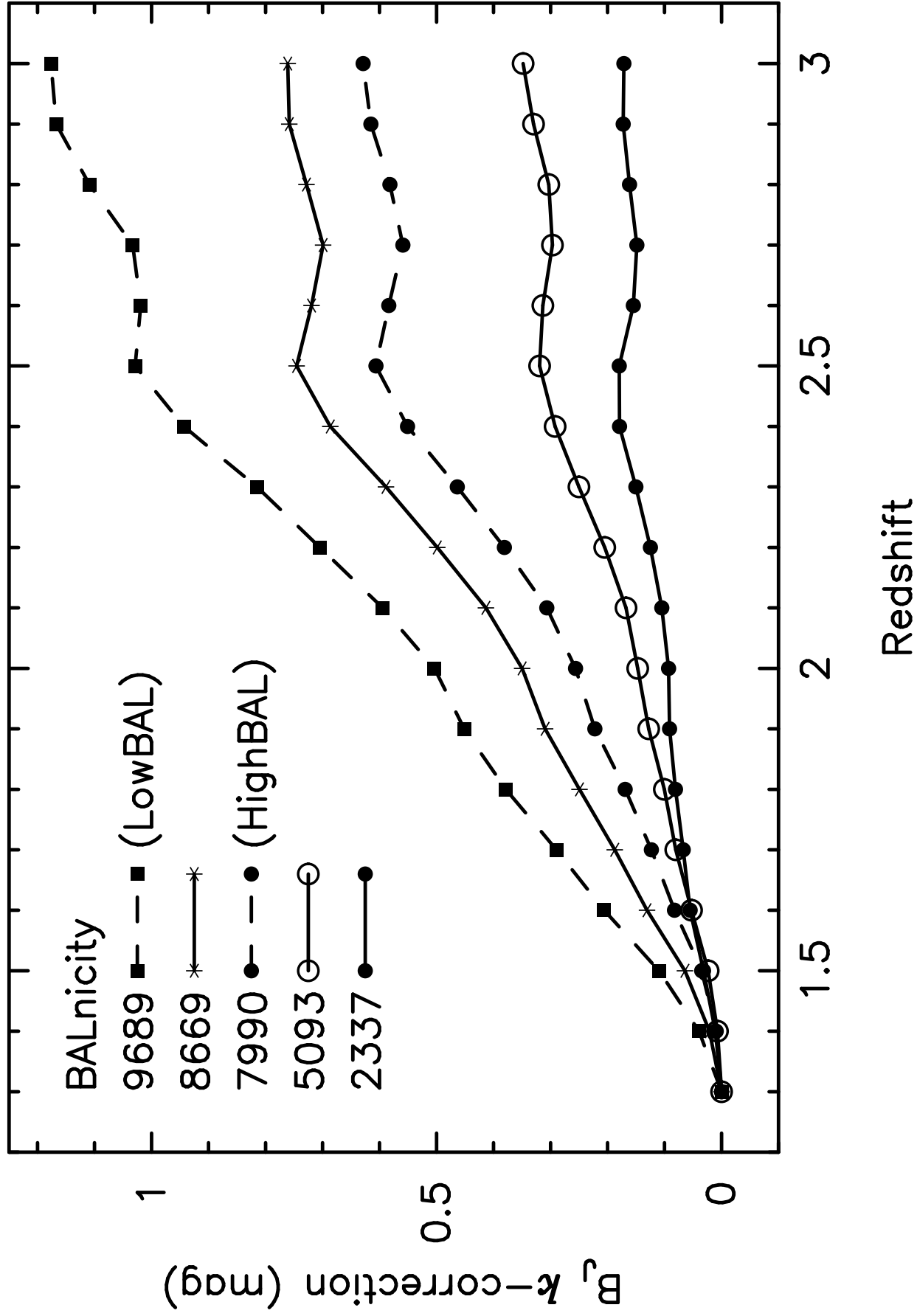
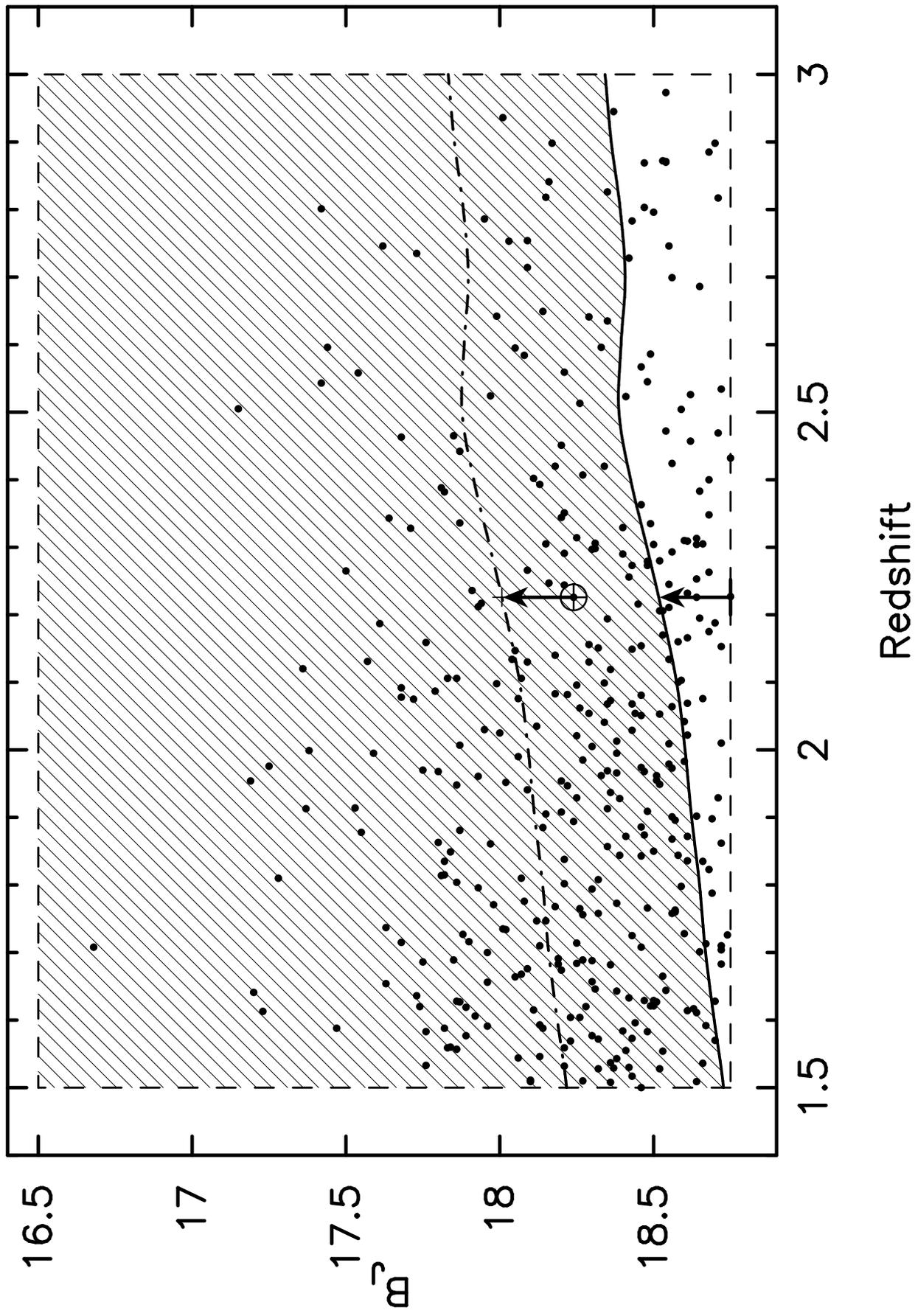
\end{document}